\documentclass[aps,prd,showpacs,notitlepage,nofootinbib,superscriptaddress,floatfix,showkeys,twocolumn]{revtex4-1}
\usepackage{float}
\usepackage{siunitx}
\usepackage{graphicx}
\usepackage{gensymb}
\usepackage{amsmath, cases}
\usepackage[usenames]{color}
\usepackage{amssymb}
\usepackage{hyperref}
\newcommand{\be}{\begin{equation}}
\newcommand{\ee}{\end{equation}}
\newcommand{\bea}{\begin{aligned}}
\newcommand{\eea}{\end{aligned}}
\newcommand{\pr}{\partial}

\newcommand{\bse}{\begin{subequations}}
\newcommand{\ese}{\end{subequations}}

\newcommand{\gv}[1]{\ensuremath{\mbox{\boldmath$ #1 $}}} 
\newcommand{\pd}[2]{\frac{\partial #1}{\partial #2}} 
\renewcommand{\v}[1]{\ensuremath{\mathbf{#1}}} 
\newcommand{\grad}[1]{\gv{\nabla} #1} 
\renewcommand{\div}[1]{\gv{\nabla} \cdot #1} 
\newcommand{\curl}[1]{\gv{\nabla} \times #1} 
\newcommand{\bmm}{\begin{multline}}
\newcommand{\emm}{\end{multline}}
\newcommand{\dagg}{\dagger}

\newcommand{\mi}{\mathrm{i}}
\begin{document}
\title{Sonofermionescence: Fermions from Ringing Bubble of Sonoluminescence}
\author{Rajesh Karmakar}
\email{rajesh018@iitg.ac.in}
\affiliation{Department of Physics, 
Indian Institute of Technology Guwahati, Assam 781039, India}
\author{Debaprasad Maity}
\email{debu@iitg.ac.in}
\affiliation{Department of Physics, 
Indian Institute of Technology Guwahati, Assam 781039, India}


\begin{abstract}
  Time-dependent gravitational background is well known as a theoretical laboratory for quantum mechanical particle production. In the present analysis, we explore such production in a time-dependent analog system. This is the follow-up of our earlier study on the Sonoluminescence phenomenon, which is modelled in terms of analog geometry coupled with the electromagnetic field exhibiting the quantum production of photons. In the same analog geometry, we studied fermion production. Although fermions have not been observed yet in the sonoluminescence experiment, we have shown here that such a system can produce a repeated flux of fermions via the parametric resonance from a quantum vacuum. Our analysis seems to suggest that in the laboratory setup, time-dependent analog systems could be an interesting playground where phenomena of quantum mechanical particle production can be observed.               
\end{abstract}
\maketitle
\section{Introduction}
 Can fermions be produced from analog system? To realize such a scenario, we take the reader through an interesting and experimentally observed phenomenon called Sonoluminescence (for review one may look at \cite{BARBER199765, Brenner:2002zz}), in which photons are produced by a yet unknown mechanism. The system consists of a quasi-periodically oscillating gas bubble in a liquid excited by sound waves \cite{Gaitan}. During its oscillation, the bubble emits flashes of light periodically \cite{Hiller:1992qz, Gompf, Weninger, Camara}. There are two competing proposals for the mechanism of such fascinating phenomena. The classical one attributes it to the well known thermal Bemisterlang processes from the high-temperature gas inside the bubble that is assumed to be partially charged \cite{BARBER199765, Barber, Lohse1999}. And the quantum mechanical one attributes it to the fact that flashes are produced from the quantum vacuum, where the quantum photon field is coupled with the gas inside the oscillating bubble that is modelled in terms of time-varying dielectric constant \cite{schwinger1992a, schwinger1992b, schwinger1993a, schwinger1993b, schwinger1993c, schwinger1993d, schwinger1994a, Liberati:1998wg, Visser:1998bqu, Liberati:1999jq, Liberati:1999uw, Eberlein:1995ex, Eberlein:1995ev}. All the proposed models have some success with their merits and demerits. For details see the review \cite{Brenner:2002zz}. We will be particularly interested in the quantum mechanical modelling of such phenomena. However, it is important to note that all the existing quantum mechanical mechanisms suffer from ultraviolet divergences \cite{Brevik:1998zs, Milton:1996wm, Milton:1997ky, Lambrecht:1996rb}, and an ad hoc cut-off is being used to accommodate the experimental observation. 
 
 In our earlier paper \cite{Karmakar:2023yce}, we have proposed a new non-perturbative mechanism, that has been shown to have the potential to explain such phenomena which does not suffer from any ultraviolet divergences. Thanks to the analog gravity interpretation of a fluid system, where dynamics of the perturbation modes can be described on an analog gravity background \cite{Unruh:1976db}. 
And in the context of sonoluminescence, we modelled the oscillating bubble in terms of an oscillating analog geometry and proposed a coupling prescription of the electromagnetic perturbation with the geometry. With the proposed coupling we indeed showed that flashes of photon are produced from quantum vacuum through parametric resonance. The geometric approach of our proposed mechanism is universal in nature.

The universality of particle production in a time-dependent background motivates us to extend our previous proposal of sonoluminescence into the fermion sector and show that along with the flashes of light, such oscillating bubbles may also produce flashes of fermions from quantum vacuum again via parametric resonance. Since the produced fermions are very light, those fermions might also be identified as neutrinos. Neutrinos are feebly interacting particles with definite chirality, and take part in weak interaction in standard model of particle physics. Neutrino oscillation experiments suggest \cite{SajjadAthar:2021prg, Giunti:2007} that they are massive with the highest mass being as high as $0.5$ eV \cite{Palanque-Delabrouille:2015pga, Planck:2015fie}, taking into account all the flavors. However, there is no experimental bound on the lowest possible neutrino mass. In our analysis, we, therefore, assume the mass of the fermion as a free parameter, however small. There are several scenarios where fermion production has been investigated, specifically in the context of early universe cosmology, in time-dependent background \cite{Dolgov:1989us, Adshead:2015jza, delRio:2014cha, Adshead:2015kza, Ema:2019yrd, Herring:2020cah, Chung:2011ck, Peloso:2000hy, Greene:1998nh, Greene:2000ew}. However, the possibility of producing fermions in the analog system is not intuitively obvious, and precisely this is the phenomenon we wish to explore.  We call this phenomenon {\it Sonofermionescence}, which may possibly be observed in the existing sonoluminescence experiment in the near future. This can be thought of as an analog cosmological laboratory.   

We have organized the paper in the following manner. In Sec.\ref{bubble_dynamics}, we begin with a brief discussion about the dynamics of the bubble in fluid and possible experimental setup for the Sonofermionescence event. This section also serves to illustrate the construction of the analog-acoustic metric in terms of the radial dynamics of the bubble. Next, in Sec.\ref{eom.ferm} we formulate the governing equations of the minimally coupled fermion field in the time-dependent background and follow the necessary steps for the canonical quantization of the classical fermion field. In Sec.\ref{num.spec} we define the suitable observable quantity, fermion number spectrum, for the experimental measurement, followed by the fermion energy flux in Sec.\ref{energyflux}, and we demonstrate the corresponding results obtained using numerical computation. Finally, we conclude with possible future scenarios.

Note that we will use the metric signature as $(-,+,+,+)$ and follow the natural units, $\hbar=c=G=1$, throughout the discussion. Also, Latin alphabets are used for the flat space time indices, while Greek alphabets are utilized to denote the curved space time indices.

\section{Sonoluminescence bubble and analog geometry}\label{bubble_dynamics}
\begin{figure}[H] 
\includegraphics[scale=0.32]{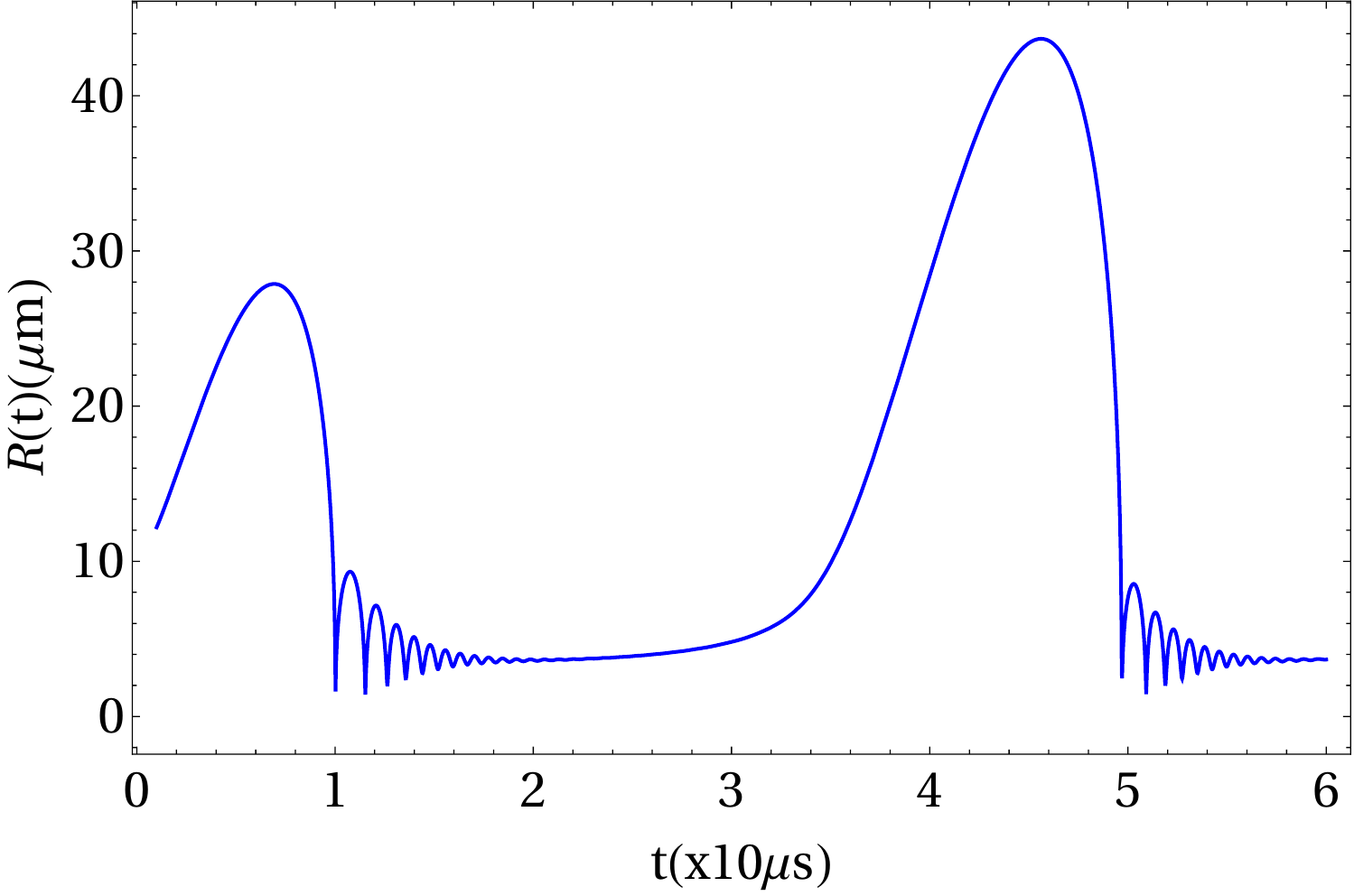}
\caption{Dynamics of the oscillating bubble is presented in terms of the temporal profile of the bubble surface. One can see that the RP equation correctly provides for the evolution of the bubble, which undergoes a repeated quasiperiodic oscillation. We have specifically shown a full period starting from $t=20{\rm \mu s}$ to $t=60{\rm \mu s}$. However, the initial fluctuations will essentially be subtracted, as we will see in the subsequent sections.}\label{radiuswt}
\end{figure}

\begin{figure}[t]
\centering
\includegraphics[scale=0.7]{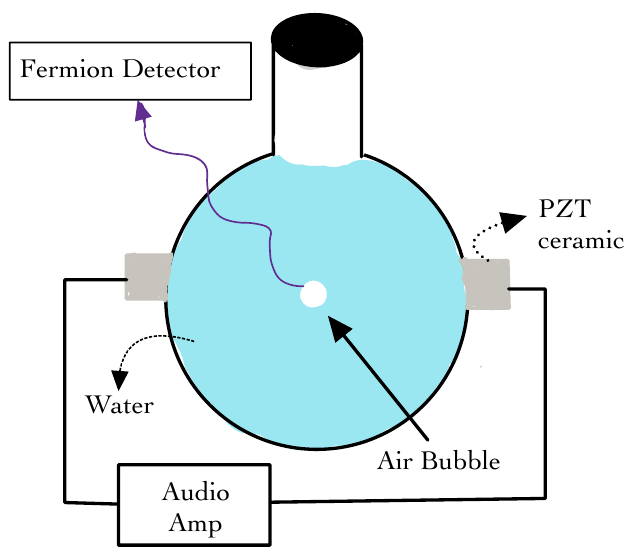}
\caption{A probable schematic of the Sonofermionoescence experimental setup, which is inspired by the Sonoluminescence experiment \cite{BARBER199765, Brenner:2002zz}.}\label{sonoexpsetup}
\end{figure}

It has been experimentally observed \cite{Gaitan, Hiller:1992qz, Gompf, Weninger, Camara} that an oscillating air bubble under the influence of the sound wave emits repeated flashes of light. Detailed experimental setup and its observational findings can be found in \cite{BARBER199765, Brenner:2002zz}. Following a similar strategy, we have proposed here the possible experimental setup, illustrated in Fig.\ref{sonoexpsetup}, for the observation of Sonofermionescence. Driven by the external sound wave, an air bubble located inside water undergoes rapid oscillation, and its dynamics is described by the well known Rayleigh-Plesset (RP) equation \cite{Barber, Rayleigh, Plesset} 
\be
\bea
&-R \ddot R \left( 1 - \frac{2 \dot R}{c_s} \right) 
-\frac{3}{2} \dot R^2 \left(1-\frac{4}{3}\frac{\dot R}{c_s}\right)\\
&+\frac{1}{\rho}\left[\frac{P_{\rm eq} R_{\rm eq}^3{\gamma}}{(R^3 - a^3)^{\gamma}} - \frac{4 \eta \dot R}{R} - \frac{2 \sigma}{R} \right]\\
&+\frac{1}{\rho c_s} \left[\frac{-3\gamma R^3 P_{\rm eq} R_{\rm eq}^{3 \gamma}}{(R^3 - a^3)^{\gamma + 1}} \dot R+ \frac{2 \sigma}{R} \dot R - \frac{4 \eta}{R} (R \ddot R - \dot R^2)\right]\\
&+ \frac{P_{\rm eq} R_{\rm eq}^{3 \gamma} \cos{\omega t}}{\rho (R^3 - a^3)^{\gamma}} - \frac{R}{\rho c_s} P_a \omega \sin{\omega t} - \frac{P_{\rm eq}}{\rho} = 0.
\eea
\ee
Where $R(t)$ denotes the bubble radius and the dot represents the time derivative. In the above equation, the value of the physical parameters, which have been experimentally used \cite{BARBER199765}, are as follows: speed of sound in water, $c_s= 1481$ m/sec, the ambient (also called equilibrium) pressure $P_{\rm eq} = 1$ atm at the ambient radius of the bubble, $R_{\rm eq} = 4.5 \mbox{ }\mu$m, the shear viscosity of the fluid, $\eta = 0.003~{\rm Kg/(m-sec)}$, coefficient of the surface tension, $\sigma =0.03~{\rm Kg/sec^2}$, the density of the fluid, $\rho\sim 1000~{\rm Kg/m^3}$, pressure-amplitude of the acoustic drive, $P_a = 1.35 $ atm, frequency of the acoustic drive, $\omega_a = 2 \pi (26.5) $ kHz. For air bubbles in water, van der Waals hard core radius is assumed to be $a = 0.5~\mbox{}\mu$m. Now, to find out the evolution of the bubble radius, we numerically solve the RP equation considering the initial conditions as $R(t=0)=4.5{\rm \mu m}$ and $\dot{R}(t =0) = 0$. We have presented the solution in Fig.\ref{radiuswt}, which depicts quasi-periodic oscillation as observed in the experiment \cite{Barber}. 
In our recent paper \cite{Karmakar:2023yce} we have modelled this dynamical bubble in terms of effective analog geometry following the prescription given in \cite{Unruh:1976db}. For the sake of completeness, we will discuss the procedure to construct this effective analog metric, which is often referred to as the acoustic metric. We start with an incompressible and irrotational fluid, and the corresponding energy-momentum tensor, $T^{\mu\nu}$ satisfies the following covariant conservation equation,
\be
\nabla_\mu T^{\mu\nu}=0,~~~T^{\mu\nu}=(\rho+P)u^\mu u^\nu+P\eta^{\mu\nu},
\ee
where $\rho$ and $P$ respectively represent the density and pressure of the fluid. $\eta^{\mu\nu}$ symbolizes the Minkowski metric depicting the flat space time. Whereas,  $u^\mu\equiv(1,\mathbf{v})$ denotes the four-velocity of the fluid. In his renowned paper \cite{Unruh:1976db}, Unruh proposed the conjecture that the behaviour of sound waves in an incompressible fluid resembles the governing equation of a classical field propagating in a gravitational background. Therefore to construct the metric one is required to derive the fluctuation equation first. For this purpose, the conservation equation above shall be represented by three coupled equations, 
\be\label{fluid.eqns}
\bea
\curl{\mathbf{v}} &= 0,\\ 
\rho\left[\frac{\pr {\mathbf{v}}}{\pr t}+ (\mathbf{v} \cdot \grad)\mathbf{v}\right] &= - \grad p - \rho \grad \Phi,\\
\frac{\pr \rho}{\pr t} + \grad \cdot (\rho \mathbf{v})& = 0. 
\eea
\ee
The first equation in the above set denotes the irrotational property of the fluid, which leads us to represent the velocity as the gradient of a velocity potential,  $\mathbf{v}=\grad \psi$. The following equations embody the Euler equation and the continuity equation, respectively. Within the Euler equation, the right-hand side indicates that the force within the medium arises from both pressure and an external driving force, quantified as the gradient of a potential, $\Phi$. To simplify our task we introduce the following parameterization as given in \cite{Unruh:1976db},
\be
g(\zeta)=\int^{e^{\zeta}}\frac{1}{\rho'}\frac{dp (\rho')}{d\rho'}d\rho',
\ee
where, $\zeta\equiv\log{\rho}$. Utilizing the above quantity, the set of fluid equations \eqref{fluid.eqns} simplify to
\be
\frac{\pr \psi}{\pr t} + \frac{1}{2}\mathbf{v}\cdot\mathbf{v} + g(\zeta) + \Phi = 0,
\ee
and
\be
\frac{\pr \zeta}{\pr t} + \mathbf{v} \cdot \grad \zeta + \div \mathbf{v}=0.
\ee
Now, one can introduce fluctuation in the medium via the fluid parameters, as follows, 
\be
\zeta=\zeta_0+\bar{\zeta},~~~\psi\to\psi_0+\bar{\psi},
\ee
where $\zeta_0$ and $\psi_0$ represent some background solutions. Then, the governing equation of the fluctuation, $\bar\psi$ turns out as \cite{Visser:1997ux, Barcelo:2005fc}, 
\be\label{eombpsi}
\bea
&\frac{1}{\rho_0} \Bigg[ \pr_t \left( \frac{\rho_0}{g'(\zeta_0)} \right) \pr_t \bar{\psi} + \pr_t\left(\frac{\rho_0{\mathbf{v}}_0}{g'(\zeta_0)}\right)\cdot\grad\bar{\psi}-\grad\cdot (\rho_0\grad\bar{\psi})\\
&~~~~~+\grad\cdot \left( \frac{\rho_0{\bf v}_0}{g'(\zeta_0)} \pr_t \bar{\psi}  \right) +\nabla \cdot \left( \mathbf{v}_0 \frac{\rho_0}{g'(\zeta_0)} \mathbf{v}_0 \cdot\grad\bar{\psi}\right)\Bigg]=0,
\eea
\ee
where,  $\mathbf{v}_0=\grad \psi_0$ and $\rho_0=e^{\zeta_0}$ are the background velocity and density of the fluid. Whereas, $g'(\zeta_0) = g'(\ln \rho_0) \approx c_0^2$, with $c_0$ representing the local velocity of sound waves within the medium. For instance, in air, $c_0\sim 343 {\rm m/s}$. Unruh proposed \cite{Unruh:1976db} that the above characteristic equation \eqref{eombpsi} of sound wave fluctuation within a fluid resembles the propagation of a field in a spacetime background possessing an effective acoustic metric (AM),  which can be expressed as,
\be \label{analog.metric}
\bea
ds_{\rm AM}^2 =&\bigg( \frac{\tilde{\rho}_0}{c^2} \bigg)^{-1}\bigg\{  \frac{(\rho_0/c^2)}{(c_0/c)^2}  \bigg[ -\big( \frac{c_0^2}{c^2}-\frac{\delta_{ij}v^i_0v^j_0}{c^2} \big)c^2 \,d{t^2}\\
&~~~~~~~~~~~~~~~~-2 \frac{v^i_0}{c}\delta_{ij}\,d{x^j} c\,dt  + \delta_{i j} \,d{x^i}\,d{x^j} \bigg]\bigg\},
\eea
\ee
where, we have introduced $\tilde{\rho}_0$ to ensure that the dimension of $ds_{\rm AM}$ corresponds to length. This new parameter can also be considered as a tuning parameter to be adjusted according to the observed results. Important to note that this overall factor results in the same fluctuation equation as \eqref{eombpsi}. Whereas $x^i$ represents the spatial coordinates and $v^i_0$ denotes the components of the local velocity of the fluid in the direction of along $x^i$. Now, we are left with the evaluation of two primary fluid parameters: the background velocity, denoted as $v_0=\sqrt{\delta_{ij}v^i_0v^j_0}$, and the background fluid density, represented by $\rho_0$. Assuming the spherically symmetric nature of the bubble oscillation, the velocity of the fluid inside the bubble, $v_0$ will be considered as radial and can approximately be taken as \cite{Barber},
\be
v_0=\frac{\dot {R}}{R}r, 
\ee
which satisfies the physical boundary conditions $v_0(r = 0) = 0$ and $v_0(r=R)=\dot R$.
Let us now consider the continuity equation for an incompressible fluid flow \cite{Visser:1997ux}, 
\be
\pd{\rho_0}{t} + \frac{\rho_0}{r^2} \pd{(r^2 v_0)}{r}=0,
\ee
Substituting the expression $v_0$, one can derive the fluid density as, 
\be
\rho_0(t)=\rho_{\rm eq}\frac{R^3_{\rm eq}}{R^3(t)},
\ee
where the equilibrium fluid density is represented as $\rho_{\rm eq}=\rho_0(t\to t_0)$, i.e. fluid density at the ambient radius of the bubble,  $R_{\rm eq}=R(t\to t_0)$. Nevertheless, with the arbitrary parameter $\tilde{\rho}_0$, we recast the expression of the term involving fluid density in front of the metric \eqref{analog.metric} as $\frac{\tilde{\rho}_0^{-1}\rho_0}{c^2_0}= \frac{\xi^3}{R(t)^3}$, where a new arbitrary parameter, $\xi$ is introduced that incorporates all the constant parameters, such as $\rho_{\rm eq}$ and $R_{\rm eq}$. Note that, $\xi$ has the dimension of length. The value of this parameter must be fixed by the observation.

In accordance with our earlier proposal \cite{Karmakar:2023yce} on photon production, we put forward the same conjecture for the fermion field, which perceives the acoustic metric as a fluctuation to the flat space time geometry. Therefore the total effective acoustic metric can be written as, 
\be
\bea
ds^2&=\bigg( - \frac{\,d{t^2}}{\epsilon} + \,d{r^2} + r^2 \,d{\Omega^2} \bigg)\\
&+  \frac{\xi^3}{R(t)^3}\bigg[ -(c_0^2 - v_0^2) \,d{t^2} - 2v_0drdt+d{r^2}+r^2d{\Omega^2}\bigg]\\
&= (g_{\mu\nu}^{(0)} + h_{\mu \nu}) \,d{x^{\mu}} \,d{x^{\nu}}.
\eea
\ee
Where we have utilized the natural unit, $c=1$, which will be followed in the remaining part of this paper unless otherwise specified. The metric described above effectively characterizes the medium within the air filled bubble. Hence, the parameter $\epsilon\sim 1$ denotes the dielectric constant of the air medium. We reexpress the above metric in a compact form in the following way, 
\be\label{eff.AM}
ds^2=-f(t,r)dt^2+2g(t,r)drdt+p(t)\left(dr^2+r^2d\Omega^2\right),
\ee
where the coefficients in the metric are,
\be\label{offdiag}
\bea
&p(t) = 1 + \frac{\xi^3}{R^3},\\
&f(t, r) =1 + \frac{\xi^3}{R^3} \left( c_0^2 - \frac{\dot{R}^2}{R^2} r^2 \right), \\
&g(t, r) = - \frac{\dot{R} \xi^3}{R^4} r.
\eea
\ee
We further simplify this effective metric by diagonalizing it with the transformation to the radial coordinate as, 
\be
\frac{\,d \bar{r}}{{1}/{p^{1/6}}} = \sqrt{p} \,d r + \frac{g}{\sqrt{p}} \,dt,
\ee
which can be integrated (for details of this derivation, see Appendix.B of \cite{Karmakar:2023yce}) to obtain the scaling in the radial coordinate, $\bar{r} = r p^{1/3}$. Utilizing this rescaling, the effective acoustic metric \eqref{eff.AM} boils down to, 
\be\label{scaled_metric}
\,d s^2 = - \left( f + \frac{g^2}{p} \right) \,dt^2 + p^{1/3} \left( \,d \bar{r}^2 + \bar{r}^2 \,d{\Omega}^2 \right).
\ee 
For the subsequent analysis, we will exclude the tilde symbol from the radial coordinate. Moreover, we will consider the fluctuation \eqref{offdiag} in the limit $r\to 0$ as done in the case of photon production from the oscillating bubble for the reason that the emission essentially happens as the bubble comes to the minimum radius \cite{Karmakar:2023yce}. Therefore the diagonal form of the metric becomes,
\be\label{scaled_metric}
\,d s^2 = -  f(t) \,dt^2 + p(t)^{\frac 1 3}\left( \,d \bar{r}^2 + \bar{r}^2 \,d{\Omega}^2 \right),
\ee
with the metric components taking the following form:
\be
\bea
p(t)=1+\frac{\xi^3}{R(t)^3}~~:~~
f(t)=1+\frac{\xi^3}{R(t)^3}c_0^2 .
\eea
\ee
Because of its spherically symmetric nature, the spatial section of the metric above can also be recast into Cartesian coordinates $(x,y,z)$, and finally we obtain,
\be\label{scaled_metric}
d s^2 = -f(t)dt^2 + p^{1/3}(t)\left(dx^2+dy^2+dz^2\right) 
\ee
In the next section, we will delve into the discussion of how this time-dependent effective acoustic metric, when coupled with a fermion field, leads to the quantum production of fermions. Note that in the computation of the produced number spectrum and flux of fermions, we have used $\xi = (0.20, 0.15, 0.10)R_{\rm eq}$ with the length scale $R_{\rm eq}$, which denotes the ambient radius of the bubble as discussed previously. For this choice of the arbitrary parameter, the condition $(\xi/R)^3<<1$ is satisfied. Therefore the fluctuation in the metric coefficients, $p(t)$ and $f(t)$ (note that $c_0$ should be put in units of the light speed, $c=3\times 10^8 {\rm m/s})$ stays in the perturbative limit. Interesting to note that the essential formalism quantitatively boils down to a conformally flat cosmological model, where quantum particle production has been explored in detail in the literature \cite{Dolgov:1989us, Adshead:2015jza, delRio:2014cha, Adshead:2015kza, Ema:2019yrd, Herring:2020cah, Chung:2011ck, Peloso:2000hy, Greene:1998nh, Greene:2000ew}. For the present case, therefore, we go to conformal coordinate, where the above metric assumes the following form,
\be\label{scaled_metric}
\,d s^2 =  p(\tau)^{\frac 1 3}(-d\tau^2 + dx^2 + dy^2 + dz^2)) ,
\ee
with conformal time $d\tau = dt\sqrt{f/p^{1/3}}$. However, as mentioned before, having considered the fluctuation perturbatively, we treat the conformal time approximately as a real time, $\tau\simeq t$. Hence, the rest of the analyses have been presented in real time coordinates.
\section{Fermions in analog background: a minimal coupling prescrition}\label{eom.ferm}
We consider the following minimally coupled action of a massive Dirac fermion $\psi$,
\be\label{action.fermion}
\mathcal{A} =\int{d^4}x\sqrt{-g} \bar{\psi} \left[i\gamma^{\mu} D_{\mu}-m_\psi \right] \psi,
\ee
where $m_\psi$ is the mass of the fermion and $g={\rm det}[g_{\mu\nu}]$, with $g_{\mu\nu}$ representing the effective acoustic metric \eqref{scaled_metric}. Curved space time gamma matrices have been denoted as $\gamma^\mu={e_a}^\mu\gamma^a$, where $\gamma^a$ represents the Flat space time gamma matrices, which satisfy the usual anti commutation relation, $\{\gamma^a,\gamma^b\}=2\eta^{ab}$ with $\eta^{ab}$ as the Minkowski metric. The connection to the curved-space gamma matrices has been made through the Vierbein, which is chosen as, ${e^a}_\mu\equiv  p^{1/6}(t){\delta^a}_\mu$. Whereas the covariant derivative can be expressed as, $D_\mu=\pr_\mu+\Gamma_\mu$, with $\Gamma_\mu=\frac{1}{4}\omega^{ab}_\mu\gamma_a\gamma_b$ \cite{Ema:2019yrd, Herring:2020cah}. From this setup we obtain $\gamma^\mu\Gamma_\mu=\frac{\dot{p}(t)}{4p^{7/6}(t)}\gamma^0$. The flat space gamma matrices, in the Dirac representation, are given by,
\be
\gamma^0=\begin{pmatrix}
I & 0 \\
0 & -I
\end{pmatrix},~~~\gamma^i=\begin{pmatrix}
0 & \sigma^i \\
-\sigma^i & 0
\end{pmatrix}.
\ee
In terms of the rescaled field $\psi(x)=p^{-1/4}(t)\chi(x)$, the equation of motion governing the fermion field can be derived from the action as, 
\be\label{1stordeom}
\mi\gamma^a\pr_a\chi-m_\psi p^{1/6}(t)\chi=0.
\ee
Conjugate momentum of $\psi$, we derived from the action as,
\be
\Pi_\chi=\frac{\pr(\sqrt{-g}\mathcal{L})}{\pr\dot{\chi}}=\mi\bar{\chi}\gamma^0, ~~~~\Pi_{\bar{\chi}}=0.
\ee
Given the flat spatial structure of the background metric, we decompose $\chi(x)$ into eigen spinors in the following way, 
{\small
\be
\bea
&\chi(x)=\sum_{\lambda= \pm 1} \int \frac{\,d^3 k}{(2 \pi)^3} \left( \hat{b}_{\v k,\lambda}\mathcal{U}_{k,\lambda}(t)+\hat{d}_{-\v k,\lambda}^{\dagg}\mathcal{V}_{-k,\lambda}(t)\right)e^{\mi\v k \cdot \v x},\\
&\chi^\dagg(x)=\sum_{\lambda= \pm 1} \int \frac{\,d^3 k}{(2 \pi)^3} \left( \hat{b}^\dagger_{\v k,\lambda}\mathcal{U}^\dagg_{k,\lambda}(t)+\hat{d}_{-\v k,\lambda}\mathcal{V}^\dagg_{-k,\lambda}(t)\right)e^{-\mi\v k \cdot \v x},
\eea
\ee}
where the creation and annihilation operator satisfies the usual anticommutation relation,
\be
\bea
&\Big\{\hat{b}_{\v k,\lambda},\hat{b}^\dagger_{\v k',\lambda'}\Big\}=(2\pi)^3\delta(\v k-\v k')\delta_{\lambda\lambda'},\\
&\Big\{\hat{d}_{\v k,\lambda},\hat{d}^\dagger_{\v k',\lambda'}\Big\}=(2\pi)^3\delta(\v k-\v k')\delta_{\lambda\lambda'}  .
\eea
\ee
Given the following quantization condition, 
\be
\{\chi_\alpha(t,\v x),\chi_\beta^\dagg(t,\v y)\}=\delta^3(\v x-\v y)\delta_{\alpha\beta},
\ee
one can show that spinor basis satisfies the following orthonormal relations \cite{Herring:2020cah},
\be\label{qcond}
\bea
&\mathcal{U}^\dagger_{k,\lambda}(t)\mathcal{U}_{k,\lambda'}(t)=\delta_{\lambda\lambda'},~~~\mathcal{V}^\dagger_{k,\lambda}(t)\mathcal{V}_{k,\lambda'}(t)=\delta_{\lambda\lambda'},\\
&\mathcal{U}^\dagger_{k,\lambda}(t)\mathcal{V}_{k,\lambda'}(t)=0,
\eea
\ee
where, 
\be
\mathcal{U}_{k,\lambda}(\tau)=
\begin{pmatrix}
f_k(t) \zeta_\lambda\\
g_k(t)\vec{\sigma}\cdot\hat{k} \zeta_\lambda
\end{pmatrix}
~;~
\mathcal{V}_{k,\lambda}=\begin{pmatrix}
-g^*_k(t) \zeta_{-\lambda}(\vec{k})\\
-f_k(t)\vec{\sigma}\cdot\hat{k} \zeta_{-\lambda}(\vec{k})
\end{pmatrix},
\ee 
with the $\zeta_\lambda$ as the eigenstates of helicity operator, $\vec{\sigma}\cdot\vec{k}\xi_\lambda=\lambda k\xi_\lambda$. The antiparticle modes can be obtained utilizing the charge conjugation operator $\mathcal{V}_{k,\lambda}=\mathcal{U}^c_{k,\lambda}\equiv {\rm C}~ \bar{\mathcal{U}}^T_{k,\lambda}=\mi \gamma^2 \mathcal{U}^*_{k,\lambda}$ and $-\mi\sigma_2\zeta^*_\lambda(\vec{k})=\lambda \zeta_{-\lambda}(\vec{k})$.
Using the normalization of the modes we derive the following relation between the time dependent function associated with the spinors as
\be\label{constr}
|f_k(t)|^2+|g_k(t)|^2=1.
\ee  
Now to find out $\mathcal{U}_{k,\lambda}(t)$ and $\mathcal{V}_{k,\lambda}(t)$ we follow the approach of \cite{Herring:2020cah}. Substituting the mode decomposition in \eqref{1stordeom} we obtain
\be\label{eomcomp}
\bea
&\mi\pr_t f_k(t)-\lambda k g_k(t)-m_\psi p^{1/6}(t)f_k(t)=0,\\
&\mi\pr_t g_k(t)-\lambda k f_k(t)+m_\psi p^{1/6}(t)g_k(t)=0.
\eea
\ee
We decouple the equation to obtain the equation of motion of the individual field as,
\be\label{2ndode}
\bea
&\pr^2_t f_k(t)+\left[k^2+m^2_\psi p^{\frac{1}{3}}(t)+\mi m_\psi\pr_t\left(p^{1/6}(t)\right) \right]f_k(t)=0,\\
&\pr^2_t g_k(t)+\left[k^2+m^2_\psi p^{\frac{1}{3}}(t)-\mi m_\psi \pr_t\left(p^{1/6}(t)\right) \right]g_k(t)=0.
\eea
\ee
To solve these differential equations numerically we consider the initial vacuum state as Bunch-Davies \cite{Birrell:1982ix, Allen:1985ux}, $f_k(t)\sim N_ke^{-\mi\omega_k t}$, with $\omega_k(t)=\sqrt{k^2+m^2_\psi p^{1/3}(t)}$ denoting the positive frequency mode, at some initial time $t=t_0$. One might consider this instant as the point when the fluctuation is turned on. The initial normalization can be fixed from \eqref{constr} as $N_k=\sqrt{(\omega_k(t)+p^{1/6}(t)m_\psi)/2\omega_k(t)}$, that will provide for the required initial conditions, $f_k(t)\sim N_ke^{-\mi\omega_k t}$ and $\pr_\tau f_k(t)\sim -\mi N_k\omega_k e^{-\mi\omega_k t}$. With this setup, we find out the solution of $f_k(t)$ from \eqref{2ndode} numerically in Mathematica with sufficient accuracy set by the accuracy goal and precision goal. This solution will be used to evaluate the observable quantities discussed in the following section.
\section{Fermion number spectrum}\label{num.spec}
\begin{figure}[h!]
\includegraphics[scale=0.33]{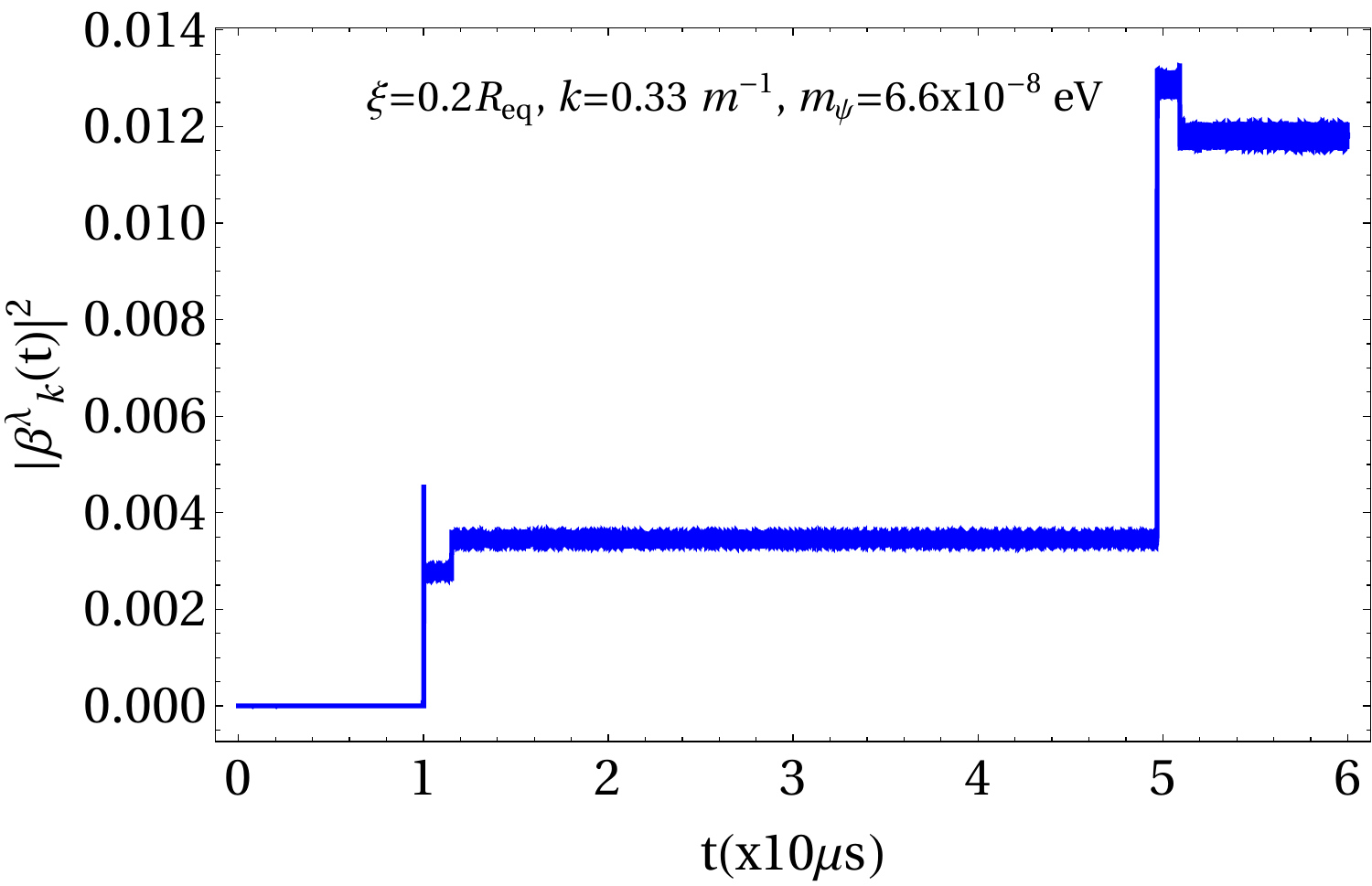}
\caption{Spectral number density has been plotted with the conformal time for fixed, $k=0.33 {\rm m^{-1}}$, tuning parameter, $\xi=0.2R_{\rm eq}$, where the description of the constant $R_0$ has been given in Sec.\ref{bubble_dynamics}. Whereas the mass of the fermion considered to be, $m_\psi=6.6\times 10^{-8}$ eV.}\label{betaksqrwtau}
\end{figure}

\begin{figure}[h!]
\includegraphics[scale=0.33]{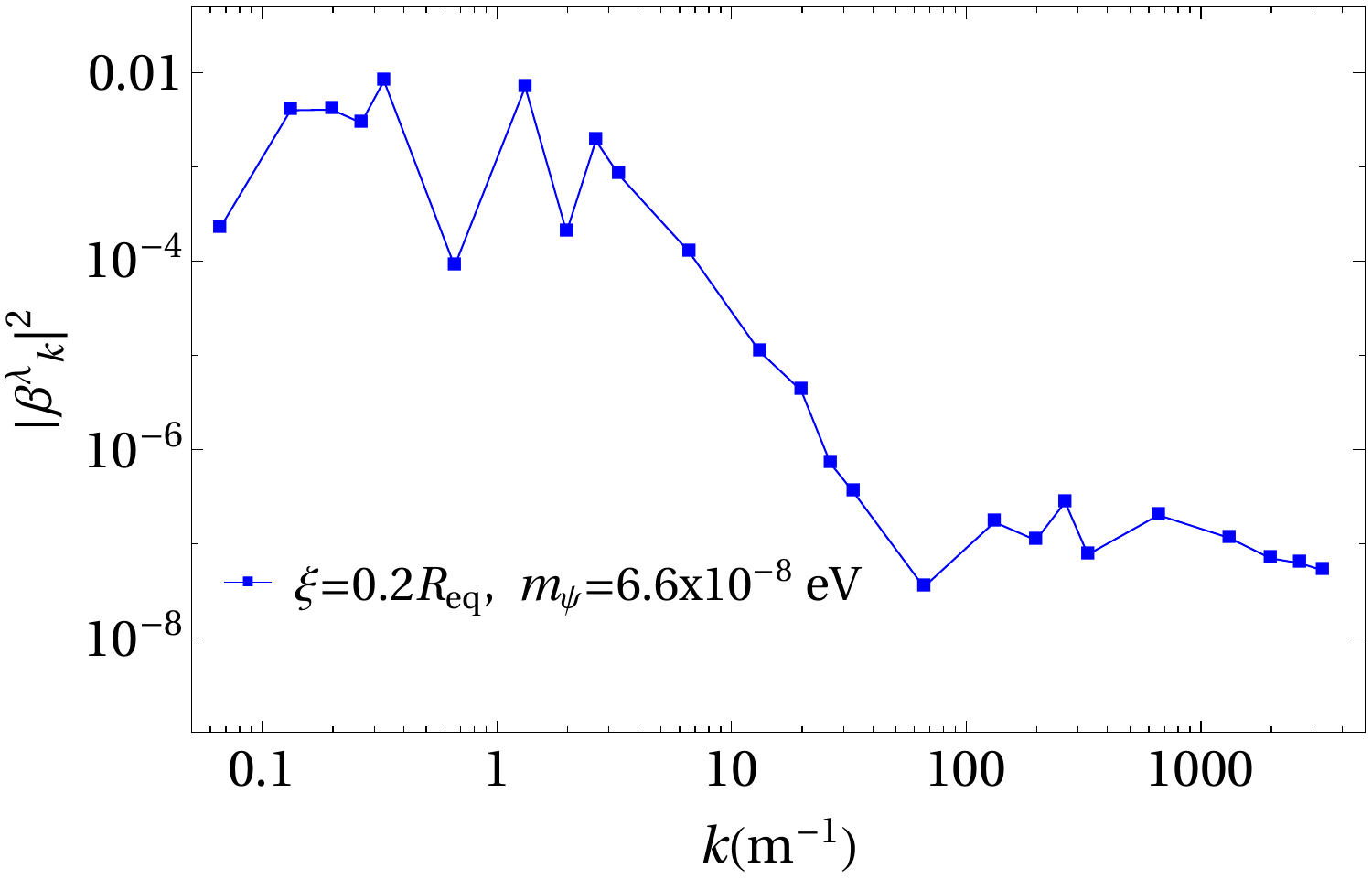}
\caption{Behaviour of the Number spectrum of the fermion. Log scale is used in both of the axes.}\label{betaksqrwk}
\end{figure}

The Hamiltonian of the system can be obtained from the action \eqref{action.fermion}, and after some simplification can be expressed as \cite{Adshead:2015kza, Peloso:2000hy}, 
\be\label{hamilton}
\bea
&H=\sum_{\lambda=\pm 1} \int \frac{\,d^3 k}{(2 \pi)^3}\Big[\left(\hat{b}^\dagger_{\v k,\lambda}\hat{b}_{\v k,\lambda}-\hat{d}_{-\v k,\lambda}\hat{d}_{-\v k,\lambda}^{\dagg}\right)E_k(t)\\
&+\hat{d}_{-\v k,\lambda}\hat{b}_{\v k,\lambda}F_k(t)+\hat{b}^\dagger_{\v k,\lambda}\hat{d}_{-\v k,\lambda}^{\dagg}F^*_k(t)\Big],\\
\eea
\ee
where, energy $E_k$ and squeezing $F_k$ functional are,
\be\label{EandF}
\bea
&E_k(t)={\rm Im}\{f_k(t)\pr_t f^*_k(t)+g_k(t)\pr_t g^*_k(t)\}\\
&~~~~~~~~=-2\lambda k {\rm Re}[f_k(t)g^*_k(t)]+m_\psi p^{1/6}(t)\{2|f_k(t)|^2-1\},\\
&F_k(t)=-\mi(-f_k(t)\pr_t g_k(t)+g_k(t)\pr_t f_k(t)).
\eea
\ee
To diagonalize the above Hamiltonian we represent the time-dependent creation and annihilation operators as follows,
\be
\bea
&\hat{b}_{\v k,\lambda}(t)=\alpha_k(t)\hat{b}_{\v k,\lambda}+\beta_k(t)\hat{d}^\dagger_{-\v k,\lambda'},\\
&\hat{d}^\dagger_{\v k,\lambda'}(t)=-\beta^*_k(t)\hat{b}_{\v k,\lambda}+\alpha^*_k(t)\hat{d}^\dagger_{-\v k,\lambda'},
\eea
\ee
with the Bogoliubov coefficients, $\alpha_k(t),\beta_k(t)$ satisfying the following constraint, 
\be\label{number.constraint}
|\alpha_k|^2+|\beta_k|^2=1.
\ee
Now the diagonalization of the Hamiltonian \eqref{hamilton} requires that the coefficient of $\hat{d}_{-\v k,\lambda}\hat{b}_{\v k,\lambda}$ to be zero \cite{Adshead:2015kza}, and can be quantitatively expressed as, 
\be
\bea
&2E_k(t)\beta^*_k(t)\alpha_k(t)+F_k(t)\alpha^2_k(t)-F^*_k(t)\beta^{*2}_k(t)=0,\\
\eea
\ee
which leads to the following expression of the Bogoliubov coefficient, 
\be
|\beta^\lambda_k(t)|^2=\frac{|F_k|^2(t)}{2\omega_k(t)(E_k(t)+\omega_k(t))}=\frac{\omega_k(t)-E_k(t)}{2\omega_k(t)}.
\ee
Note that we have used $E^2_k(t)+F^2_k(t)=\omega^2_k(t)$ to derive the last equality. Finally, substituting the expression of $E_k(t)$ from \eqref{EandF} in the above equation we formulate the occupation number density $|\beta^\lambda_k|^2$ as \cite{Peloso:2000hy, Ema:2019yrd},
\be
\bea  \label{beta}
|\beta^\lambda_k(t)|^2
=\frac{1}{2}+\frac{m_\psi p^{1/6}-2i(f^*_k \pr_t f_k-f_k \pr_t f_k^*)}{2\omega_k(t)},
\eea
\ee
We now analyze in detail the characteristics of the fermion number spectrum $|\beta^\lambda_k(t)|^2$ in terms of tuning parameters $\xi$ and fermion mass, $m_\psi$. After numerically solving the Dirac equation \eqref{eomcomp}, and substituting the solution in \eqref{beta}, one will be able to evaluate the number density of the fermion. We have plotted the time evolution of the fermion number density $|\beta^\lambda_k(t)|^2$ in Fig.\ref{betaksqrwtau} for a fixed wave number $k=0.33 {\rm m^{-1}}$ and mass $m_\psi= 6.6\times 10^{-8}$ eV of the fermion. From the plot, it is clear that at the time when the bubble shrinks to its minimum radius say $t=60 {\rm \mu s}$ (see Fig.\ref{radiuswt}), the sudden jump in the fermion number density occurs, and such a jump is repeated with the same time period as that of the oscillating bubble. Therefore, the parametric resonance, which is responsible for the photon flux as observed in the experiment \cite{Gaitan, Hiller:1992qz, Gompf}, also causes the production of flashes of fermions and can in principle be detected. We further plotted the number spectrum by taking the increment in photon number density from $t=45{\rm \mu s}$ to $t=55 {\rm \mu s}$,  which shows a peak near around $1~\mbox{m}^{-1}$. We have been able to obtain a stable solution within the range $(0.01, 10^5)~\mbox{m}^{-1}$ of $k$, as shown in blue dotted points Fig.\ref{betaksqrwk}. Further numerical investigation is needed to extend that range to a higher frequency. We should also mention that the magnitude of the produced number density is small compared to the maximum allowable value of $|\beta_k|^2 \sim 1$ \eqref{number.constraint}. It will be interesting to consider explicit conformal breaking coupling as has been utilized for photon \cite{Karmakar:2023yce} and check whether this magnitude could be enhanced. We save this analysis as our future project.
In the next section, we define fermion energy flux that can also be experimentally measured.  

\section{Fermion Energy Flux}\label{energyflux}
\begin{figure}[t]
\includegraphics[scale=0.33]{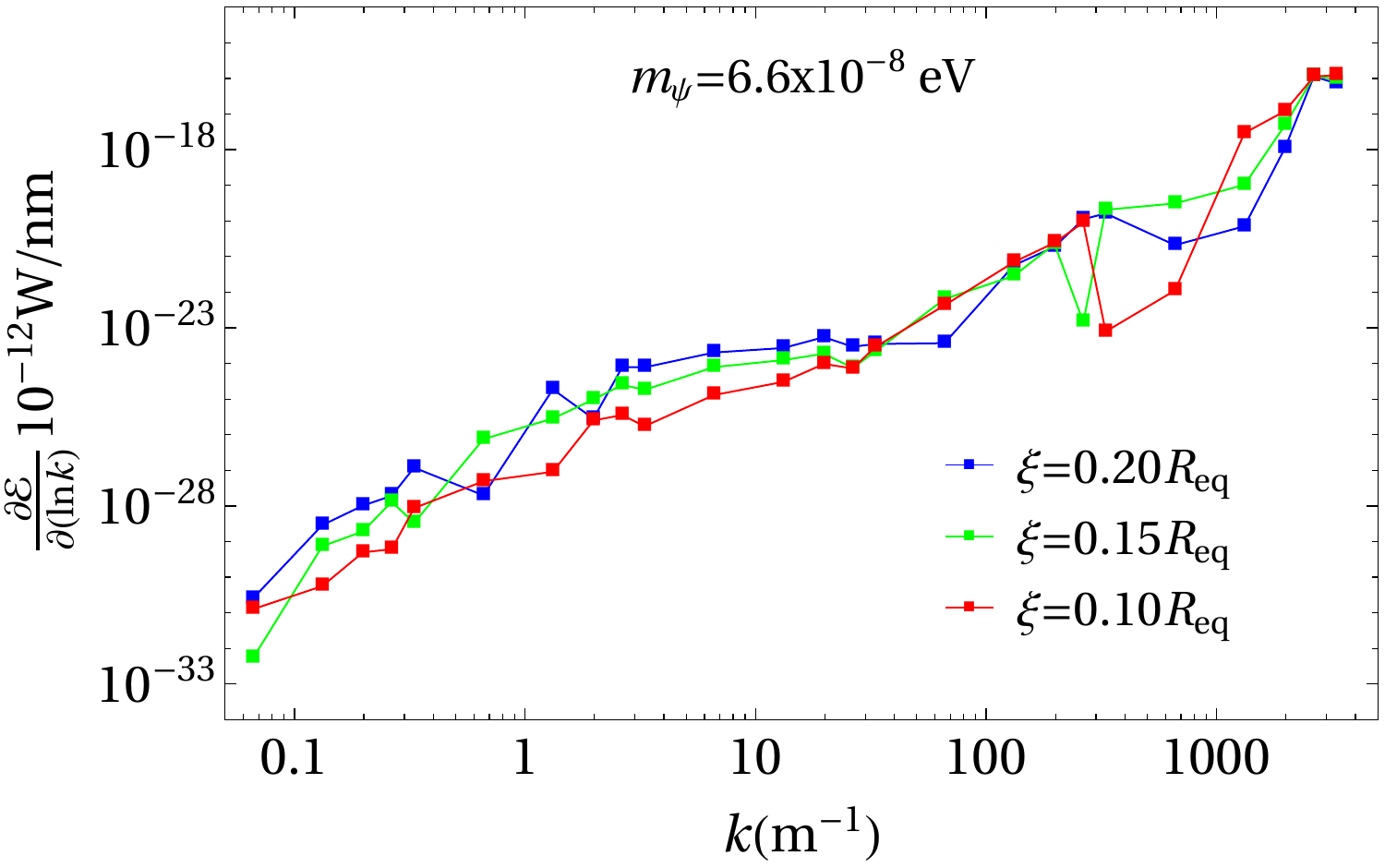}
\caption{Fermion energy flux per unit time has been plotted with the frequency, $k$ for various values of the tuning parameter $\xi$.}\label{flux}
\end{figure}

\begin{figure}[t]
\includegraphics[scale=0.33]{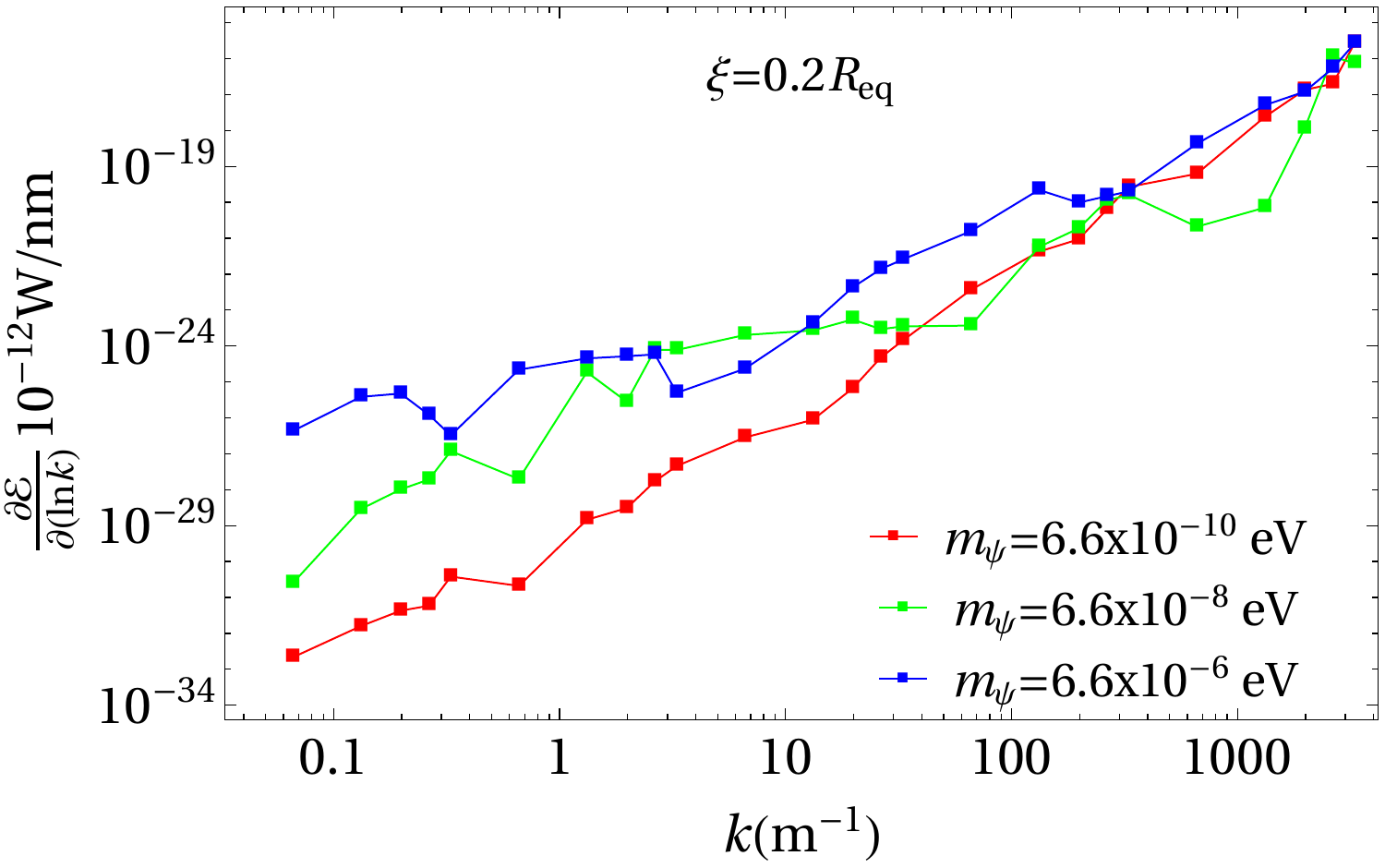}
\caption{Fermion energy flux per unit time has been plotted with the frequency, $k$ for the different mass of the fermion. We have also kept the background parameter, $\xi$, fixed.}\label{massvar}
\end{figure}

The expectation value of this Hamiltonian can be expressed in terms of fermion number spectrum as \cite{Ema:2019yrd},
\be
\langle H\rangle=\sum_{\lambda=1,2}\int\frac{d^3k}{(2\pi)^3}~\omega_k(t)|\beta^\lambda_k(t)|^2\delta^3(0),
\ee
where $\delta^3(0) = \mbox{spatial volume}$ \cite{Mukhanov:2007zz}. This immediately leads to spectral energy density as,
\be
\frac{\pr}{\pr\ln k}\left( \frac{\langle H\rangle}{\delta^3(0)}\right)=\sum_{\lambda=1,2}~4\pi k^3\omega_k(t)|\beta^\lambda_k(t)|^2.
\ee
The dimension of the above quantity is simply energy per unit volume. Upon converting into the unit of Watt per nanometer which is convenient for the experimental measurement we have 
\be
\frac{\mbox{Flux}}{s \times \mbox{nm}} = \frac{\pr\mathcal{E}}{\pr\ln k} \times \frac{4\pi(0.2\mu\mbox{m})^2}{50{\rm ps}} \left[\frac{\mbox{Watt}}{\mbox{nm}}\right],
\ee
where, in the definition, we have made use of the experimentally measured bubble radius $\sim 0.2{\rm \mu m}$\cite{Camara} at the moment of emission of the photon flux with pulse width $\sim 50{\rm ps}$ \cite{Hiller:1992qz, Weninger}, as fermion emission will follow the same pattern as that of the photons. We numerically evaluate the above expression using  \eqref{betaksqrwtau}, and plotted in Fig.\ref{flux} for three different values of the parameter $\xi = (0.20, 0.15, 0.10)R_{\rm eq}$, with the wave number $k$. Interestingly, one can see the increasing trend in the flux with the wave number $k$. As expected, from the coupling prescription with the background, increasing the value of $\xi$ will lead to more production, which is clearly visible from Fig.\ref{flux} in the low frequency range up to $k\sim 100 {\rm m^{-1}}$. 
However, in the high frequency range, generally, production being very less (see fig.\ref{betaksqrwk}) and frequency of the modes playing the dominant role, we find it hard to distinguish the effect of $\xi$. Now, recall that the proposed analog metric \eqref{scaled_metric} is conformally flat, hence the fermion mass plays the primary role in realising the parametric resonance as it is the mass term \eqref{1stordeom} which breaks the conformal invariance. Therefore, we have further plotted the energy flux in Fig.\ref{massvar} for three different values of fermion mass, $m_\psi = (6.6 \times 10^{-10}, 6.6 \times 10^{-8}, 6.6 \times 10^{-6})$ eV. This mass range could be very well suited for identifying the fermions as neutrinos. Initially, the energy spectrum increases in magnitude as we increase the mass, and then it saturates at around $m_\psi \sim 10^{-6}$ eV. Saturation is more explicit in the high frequency range as the frequency plays the dominant role in the effective potential of the equation of motion \eqref{2ndode} at a higher frequency. The sudden jump of the bubble during moments of transition to the minimum radius poses a challenge in obtaining the numerical solution for heavier masses, which appears in the equation of motion \eqref{2ndode} as a multiplying factor with the metric coefficients. Hence, a more advanced numerical method (such as the methods discussed in \cite{Elrod:2022, Hairer:1999}) suitable to handle stiff ordinary differential equations is required to present a better representation of the spectrum for heavier mass. Nevertheless, in the present analysis, we have mainly shown that the fermion can be produced periodically via parametric amplification \cite{Greene:1998nh}. To our knowledge, this is the first case study of fermion production in the context of analog fluid medium.
\section{Conclusion} 
In the time dependent background quantum particle production is an intriguing phenomena which is the subject of investigation for a long time \cite{Dolgov:1989us, Adshead:2015jza, delRio:2014cha, Adshead:2015kza, Ema:2019yrd, Herring:2020cah, Chung:2011ck, Peloso:2000hy, Greene:1998nh, Greene:2000ew}. It has been successfully applied in the early universe cosmological scenario. It would indeed be interesting if such phenomena could be observed in real laboratory systems. Sonoluminescence is assumed to be one such phenomena where quantum photon production is believed to be one of the possible mechanisms. In our recent proposal \cite{Karmakar:2023yce} we have revisited this issue and proposed a formalism in the framework of analog geometric model. We further emphasised that non-perturbative parametric resonance plays a central role in explaining such phenomena.

The geometric framework is universal in nature which motivates us to look into the possibility of fermion production in such an analog model. In this article, we propose a minimal coupling prescription of the fermion field with the analog oscillating bubble geometric. Along with the flashes of photons, the fermion field is indeed observed to experience parametric amplification due to the oscillating bubble, and that produces flashes of fermion from the quantum vacuum. 

We have computed both the fermion number and energy density spectrum in the experimental unit. We have obtained the flux in the low frequency region up to $\sim 10^5 {\rm m}^{-1}$. Interestingly the produced flux in terms of frequency $(k)$ can be matched with the experimental results if it is observed. The smallness of the magnitude can be attributed to the fact that we have restricted our analysis to purely minimal interaction with the background acoustic metric. To enhance production, further exploration with more explicit conformal breaking coupling could be an interesting project for the future.

\noindent
\textbf{Acknowledgments:} We thank our
Gravity and High Energy Physics groups at IIT Guwahati for useful discussions and suggestions. The initial collaboration with Chandramouli Chowdhury for this project is also gratefully acknowledged. RK wants to thank Bibhas Ranjan Majhi for the insightful discussion on the observer-dependent aspects of particle production in curved space time.

\textbf{Data availability:} This manuscript has no associated data added to any data repository. No data are associated in this manuscript.


\begin{thebibliography}{99}
\bibitem{BARBER199765}
B~P.~Barber, R.A.~Hiller, R~Löfstedt, S.J.~Putterman, and K.R.~Weninger,
Physics Reports. 281 (2): 65–143
\bibitem{Brenner:2002zz}
M.~P.~Brenner, S.~Hilgenfeldt and D.~Lohse,
Rev. Mod. Phys. \textbf{74}, 425-484 (2002)
\bibitem{Gaitan}
D. F. Gaitan, L. A. Crum, C. C. Church, and R. A. Roy,
J. Acoust. Soc. Am. 91, 3166–3183 (1992)
\bibitem{Hiller:1992qz}
R.~Hiller, S.~J.~Putterman and B.~P.~Barber,
Phys. Rev. Lett. \textbf{69}, 1182-1184 (1992)
\bibitem{Gompf}
B. Gompf, R. Günther, G. Nick, R. Pecha, and W. Eisenmenger
Phys. Rev. Lett. 79, 1405
\bibitem{Weninger}
R.A.~Hiller, S.J.~Putterman, and K.R.~Weninger,
Phys. Rev. Lett. 80, 1090
\bibitem{Camara}
Carlos Camara, Seth Putterman, and Emil Kirilov,
Phys. Rev. Lett. 92, 124301
\bibitem{Barber}
R.~Löfstedt, B.~P.~Barber, and S.~J.~Putterman 
Physics of Fluids A: Fluid Dynamics 5, 2911 (1993); 
\bibitem{Lohse1999}
S.~Hilgenfeldt, S.~ Grossmann, D.~Lohse (1999). 
Nature. 398. 402-405. 10.1038/18842. 
\bibitem{schwinger1992a}
J. Schwinger, Proc. Nat. Acad. Sci. 89, 4091–4093 (1992).
\bibitem{schwinger1992b}
J. Schwinger, Proc. Nat. Acad. Sci. 89, 11118–11120 (1992).
\bibitem{schwinger1993a}
J. Schwinger, Proc. Nat. Acad. Sci. 90, 958–959 (1993).
\bibitem{schwinger1993b}
J. Schwinger, Proc. Nat. Acad. Sci. 90, 2105–2106 (1993).
\bibitem{schwinger1993c}
J. Schwinger, Proc. Nat. Acad. Sci. 90, 4505–4507 (1993).
\bibitem{schwinger1993d}
J. Schwinger, Proc. Nat. Acad. Sci. 90, 7285–7287 (1993).
\bibitem{schwinger1994a}
J. Schwinger, Proc. Nat. Acad. Sci. 91, 6473–6475 (1994).
\bibitem{Liberati:1998wg}
S.~Liberati, F.~Belgiorno, M.~Visser and D.~W.~Sciama,
J. Phys. A \textbf{33}, 2251-2272 (2000)
\bibitem{Visser:1998bqu}
M.~Visser, S.~Liberati, F.~Belgiorno and D.~W.~Sciama,
Phys. Rev. Lett. \textbf{83}, 678-681 (1999)
\bibitem{Liberati:1999jq}
S.~Liberati, M.~Visser, F.~Belgiorno and D.~Sciama,
Phys. Rev. D \textbf{61}, 085023 (2000)
\bibitem{Liberati:1999uw}
S.~Liberati, M.~Visser, F.~Belgiorno and D.~Sciama,
Phys. Rev. D \textbf{61}, 085024 (2000)
\bibitem{Eberlein:1995ex}
C.~Eberlein,
Phys. Rev. Lett. \textbf{76}, 3842-3845 (1996)
\bibitem{Eberlein:1995ev}
C.~Eberlein,
Phys. Rev. A \textbf{53}, 2772-2787 (1996)
\bibitem{Brevik:1998zs}
I.~H.~Brevik, V.~N.~Marachevsky and K.~A.~Milton,
Phys. Rev. Lett. \textbf{82}, 3948-3951 (1999)
\bibitem{Milton:1996wm}
K.~A.~Milton and Y.~J.~Ng,
Phys. Rev. E \textbf{55}, 4207-4216 (1997)
\bibitem{Milton:1997ky}
K.~A.~Milton and Y.~J.~Ng,
Phys. Rev. E \textbf{57}, 5504-5510 (1998)
\bibitem{Lambrecht:1996rb}
A.~Lambrecht, M.~T.~Jaekel and S.~Reynaud,
Phys. Rev. Lett. \textbf{78}, 2267 (1997)
\bibitem{Karmakar:2023yce}
R.~Karmakar and D.~Maity,
Phys. Rev. D \textbf{109}, no.10, 105016 (2024)
\bibitem{Unruh:1976db}
W.~G.~Unruh,
Phys. Rev. D \textbf{14}, 870 (1976)
\bibitem{SajjadAthar:2021prg}
M.~Sajjad Athar, S.~W.~Barwick, T.~Brunner, J.~Cao, M.~Danilov, K.~Inoue, T.~Kajita, M.~Kowalski, M.~Lindner and K.~R.~Long, \textit{et al.}
Prog. Part. Nucl. Phys. \textbf{124}, 103947 (2022)
\bibitem{Giunti:2007}
 Giunti, Carlo, and Chung W. Kim, Fundamentals of Neutrino Physics and Astrophysics (Oxford, 2007; online edn, Oxford Academic, 1 Jan. 2010)
\bibitem{Palanque-Delabrouille:2015pga}
N.~Palanque-Delabrouille, C.~Y\`eche, J.~Baur, C.~Magneville, G.~Rossi, J.~Lesgourgues, A.~Borde, E.~Burtin, J.~M.~LeGoff and J.~Rich, \textit{et al.}
JCAP \textbf{11}, 011 (2015)
\bibitem{Planck:2015fie}
P.~A.~R.~Ade \textit{et al.} [Planck],
Astron. Astrophys. \textbf{594}, A13 (2016)
\bibitem{Dolgov:1989us}
A.~D.~Dolgov and D.~P.~Kirilova,
Sov. J. Nucl. Phys. \textbf{51}, 172-177 (1990)
JINR-E2-89-321.
\bibitem{Adshead:2015jza}
P.~Adshead and E.~I.~Sfakianakis,
Phys. Rev. Lett. \textbf{116}, no.9, 091301 (2016)
\bibitem{delRio:2014cha}
A.~del Rio, J.~Navarro-Salas and F.~Torrenti,
Phys. Rev. D \textbf{90}, no.8, 084017 (2014)
\bibitem{Adshead:2015kza}
P.~Adshead and E.~I.~Sfakianakis,
JCAP \textbf{11}, 021 (2015)
\bibitem{Ema:2019yrd}
Y.~Ema, K.~Nakayama and Y.~Tang,
JHEP \textbf{07}, 060 (2019)
\bibitem{Herring:2020cah}
N.~Herring and D.~Boyanovsky,
Phys. Rev. D \textbf{101}, no.12, 123522 (2020)
\bibitem{Chung:2011ck}
D.~J.~H.~Chung, L.~L.~Everett, H.~Yoo and P.~Zhou,
Phys. Lett. B \textbf{712}, 147-154 (2012)
\bibitem{Peloso:2000hy}
M.~Peloso and L.~Sorbo,
JHEP \textbf{05}, 016 (2000)
\bibitem{Greene:1998nh}
P.~B.~Greene and L.~Kofman,
Phys. Lett. B \textbf{448}, 6-12 (1999)
\bibitem{Greene:2000ew}
P.~B.~Greene and L.~Kofman,
Phys. Rev. D \textbf{62}, 123516 (2000)
\bibitem{Rayleigh}
Lord Rayleigh,
Philos. Mag. 34, 94 (1917);
\bibitem{Plesset}
M. Plesset,
J. Appl. Mech. 16, 277 (1949)
\bibitem{Visser:1997ux}
M.~Visser,
Class. Quant. Grav. \textbf{15}, 1767-1791 (1998)
\bibitem{Barcelo:2005fc}
C.~Barcelo, S.~Liberati and M.~Visser,
Living Rev. Rel. \textbf{8}, 12 (2005)
\bibitem{Birrell:1982ix}
N.~D.~Birrell and P.~C.~W.~Davies,
Cambridge Univ. Press, 1984,
\bibitem{Allen:1985ux}
B.~Allen,
Phys. Rev. D \textbf{32}, 3136 (1985)
\bibitem{Mukhanov:2007zz}
V.~Mukhanov and S.~Winitzki,
Cambridge University Press, 2007,
ISBN 978-0-521-86834-1, 978-1-139-78594-5
\bibitem{Elrod:2022}
Elrod, C., Ma, Y., Althaus, K., \& Rackauckas, C. (2022, September).
In 2022 IEEE High Performance Extreme Computing Conference (HPEC) (pp. 1-9). IEEE.
\bibitem{Hairer:1999}
Hairer, Ernst, and Gerhard Wanner. ``Stiff differential equations solved by Radau methods." Journal of Computational and Applied Mathematics 111.1-2 (1999): 93-111.
\end{thebibliography}
\end{document}